\input harvmac  
\input tables 
\thicksize=1.5pt
\noblackbox 
\def\lfm{\smallskip\noindent\item} 
 
\Title{hep-ph/9607362, UW/PT 96-09, RU-96-59} 
{\vbox{ \centerline{A
Realistic Supersymmetric Model with} 
\medskip \centerline{Composite Quarks} }}
\bigskip \centerline{ Ann E. Nelson} 
\smallskip \centerline{\it Department of Physics, Box 351560} 
\centerline{\it University of Washington, Seattle, WA 98195-1560} 
\medskip \centerline{Matthew J. Strassler} 
\smallskip 
\centerline{ \it Department of Physics and Astronomy, Rutgers University} 
\centerline{ \it P.O. Box 849, Piscataway, NJ 08855} 
\bigskip \centerline{\bf Abstract} 
\smallskip
\noindent  We describe a realistic, renormalizable, supersymmetric
``quindecuplet'' model in which the top quark, left-handed bottom
quark, and up-type Higgs boson are composite, with a compositeness
scale $\sim 1-3$ TeV. The top-Higgs Yukawa coupling is a dynamically
generated strong interaction effect, and is naturally much larger than
any other Yukawa coupling. The light quark doublets and right-handed
up-type quarks are also composite but at higher energies; the
hierarchy of quark masses and mixings is due to a hierarchy in the
compositeness scales.  Flavor changing neutral currents are naturally
suppressed, as is baryon number violation by Planck-scale dimension
five operators.  The model predicts that the most easily observable
effects would be on $b$-quark physics and on the $\rho$ parameter.  In
particular a small negative $\Delta\rho=-\epsilon$ leads to $\Delta
R_b> +2\epsilon$. There are effects on $B$ meson mixing
and on flavor-changing neutral-current $b$-quark decays to leptons 
which might be detectable, but not on $b\rightarrow s\gamma$.  The
model also suggests the supersymmetry-breaking mass for the right handed
top squark might be considerably larger than that of the left handed
top squark.

\Date{7/96}
\newsec{Introduction}

One of the most intriguing clues to physics beyond the standard model
is the hierarchy of quark and lepton masses. In understanding why most
fermions are so much lighter than the top quark and the $W^\pm$ and
$Z^0$ gauge bosons, and why they seem to have a definite though ragged
generational structure, we might hope to learn the mechanism of
electroweak supersymmetry breaking, explain why only three generations
exist, learn where Yukawa couplings come from, and obtain hints about
GUT and Planck scale physics.

Supersymmetry provides an attractive solution to the gauge hierarchy
problem, but so far has not given us an explanation for the hierarchy
of quark and lepton Yukawa couplings.  In many supersymmetric models
the large top quark Yukawa coupling provides the dynamics behind
electroweak symmetry breaking
\ref\topbreaks{L. Ibanez and G.G. Ross, Phys. Lett. {\bf B110} (1982) 215;
 L. Alvarez-Gaume, M. Claudson and M. B. Wise, Nucl. Phys. {\bf B207} 
(1982) 96; K. Inoue et al., Prog. Theor. Phys. {\bf 68} (1982) 927;
H.P. Nilles, M. Srednicki and D. Wyler, Phys. Lett. {\bf B120} (1983) 346;
J. Ellis et al., Phys. Lett. {\bf B125} (1983) 275; L. Alvarez-Gaume,
J. Polchinski and M.B. Wise, Nucl. Phys. {\bf B221} (1983) 495}, but no
explanation is given for why this Yukawa coupling is so much larger
than the others.

A proposal along these lines was made in \ref\topYukawa{A.E. Nelson,
unpublished} and in \ref\highscale{M.J. Strassler, Phys. Lett.  {\bf
B376} (1996) 119, hep-ph/9510342} in which a dynamical mechanism for
generating the top quark mass was suggested.  In this ``quindecuplet''
scenario, the top quark, left-handed bottom quark, and up-type Higgs
are part of a 15 dimensional multiplet of composite particles, each
containing two ``preons''. The ordinary $SU(3)_c\times SU(2)_w\times
U(1)_Y$ gauge interactions can be embedded into $SU(5)$, under which
the composite particles transform as $ {\bf 5}+ {\bf 10}$.  The top
quark Yukawa coupling is generated by a strong coupling effect of
confinement \ref\seiberg{N. Seiberg, Phys. Rev. {\bf D49} (1994) 6857,
hep-th/9402044}, and the bottom quark mass is generated through an
effective higher-dimension operator. Viable three-generation models,
employing all or part of this mechanism with the compositeness scale
near to the Planck scale, were proposed in \highscale.  However, these
models are very difficult to rule out as they have no new consequences
at low energy.  Furthermore, the compositeness scale cannot be scaled
down to low energy as proton decay will become far too rapid.

 In this letter we discuss a significantly modified version of the
supersymmetric quindecuplet scenario in which the scale of
compositeness of the left handed top and bottom quarks, the right
handed top quark, and the up-type Higgs superfield can be only
slightly above the weak scale, and the proton is stable. The other
left-handed quark doublets and right-handed up and charm quarks are
similarly composite, but are made of different preons, and are much
more tightly bound. The right handed down-type quarks, the leptons,
and the down-type Higgs are elementary particles.  A hierarchy of
quark masses and mixings with a reasonable structure can be
generated. Our model provides a realization of 't Hooft's idea that
the Higgs should be composite at a scale below a few TeV and that some
of the observed fermions should be composites which, due to chiral
symmetry, are relatively light compared with their inverse size, with
the Yukawa couplings generated via compositeness effects
\ref\comp{G. 't Hooft, 1979 Cargese Lectures, published in Recent
Developments In Gauge Theories, Proceedings, Nato Advanced Study
Institute New York, Usa: Plenum (1980); see also S. Dimopoulos,
S. Raby and L. Susskind, Nucl. Phys. {\bf B173} (1980) 208}. We have taken
advantage of the recent discovery that the low energy limit of many
strongly coupled supersymmetric gauge theories contains massless
composite bound states \seiberg, (as has been anticipated for some
time, \ref\cernrep{D. Amati, K. Konishi, Y. Meurice, G.C. Rossi,
Phys. Rept. {\bf 162} (1988) 169},
\ref\susycomp{Examples of models of supersymmetric compositeness may be found
in the following: R.R. Volkas and G.C. Joshi, Phys. Rept. {\bf 159}
(1988) 303, and references therein; R.D. Peccei {\sl The Mass Problem
and the Issue of Compositeness}, Lectures given at Lake Louise Winter
Institute: Selected Topics in Electroweak Interactions, Lake Louise,
Canada, 1987; M. Yasue, Prog. Theor. Phys. {\bf 78} (1987) 1437;
S.Yu. Khlebnikov and R.D. Peccei, Phys. Rev. {\bf D48} (1993) 361,
hep-ph/9212275; J.C. Pati, {\sl Toward a Pure Gauge Origin of the
Fundamental Forces: Unity Through Preons.}  Lectures given at the Puri
Winter School in Physics: Particle Physics and Cosmology at the
Interface, Uri, India, 1993; H. Terazawa, Mod. Phys. Lett. {\bf A10}
(1995) 199; P. Dimopoulos, G.K. Leontaris and N.D. Tracas IOA-01-96,
hep-ph/9604265 }).

There is a vast literature on composite models of quarks and leptons,
with and without supersymmetry \susycomp.  However, we believe this
example is unique in having the following features:

\lfm{1.} The dynamics of the strongly coupled gauge theory we consider
  is tightly constrained by consistency with supersymmetry.
\lfm{2.} The theory is renormalizable and weakly coupled at high energy.
\lfm{3.} Many features of the hierarchy of quark masses and mixing
  angles may be qualitatively understood in terms of 3 different
  compositeness scales.  
\lfm{4.} At the weak scale, the model is a phenomenologically viable
  and interesting extension of the standard model, with new strong
  gauge interactions at 1-3 TeV. Baryon and lepton numbers are
  sufficiently conserved and new
  sources of flavor changing neutral currents (FCNC) can be kept within
  experimental bounds.

\noindent These features make the model an ideal laboratory to study the
 observable effects which could arise from compositeness.

The low energy phenomenology of the model is similar to that of the
minimal supersymmetric standard model (MSSM) \ref\mssmrev{For a review
see H. E. Haber and G.L. Kane, Phys. Rept. {\bf 117} (1985) 75}, but it has an
approximate $SU(6)$ global symmetry, of which the standard model gauge
group is a subgroup, that has several interesting consequences.
First, certain important low energy signals of compositeness,
including corrections to top quark and left-handed bottom quark
couplings and to the $\rho$ parameter, are related by $SU(6)$ and
supersymmetry.  It is amusing to note that this scenario, in which the
up-type Higgs and left-handed bottom quark are composite through the
same dynamics, can potentially explain the reported excess in
$Z\rightarrow b\bar b$ events\foot{There is also a reported (less
significant) deficit in $Z\rightarrow c \bar c$ events
\ref\rbanom{The LEP Collaborations ALEPH, DELPHI, L3, OPAL and 
the LEP Electroweak Working Group, CERN-PRE/95-172; P.B. Renton,
plenary talk at the International Conference on High Energy Physics,
Beijing (1995), LEP Electroweak Working Group and SLD Heavy Flavor
Group LEPEWWG/96-01, SLD physics note 47 (1996)}, which we cannot
account for.  A recent analysis suggests that a revision in
charmed-meson branching fractions could account for the charm deficit
in $Z$ decay, and perhaps also affect the extraction of the
$Z\rightarrow b\bar b$ rate
\ref\distw{I. Dunietz et al., Fermilab-pub-96/26-T, hep-ph/9606327}.}
\rbanom\ and push the $\rho$ parameter slightly negative without
leading to other phenomenological problems, as we show in section~4.

Below the confinement scale the $SU(6)$ symmetry requires two massive
supermultiplets which are not part of the MSSM --- a charge 1/3 color
triplet $D$ with baryon number $-2/3$ and a charge 1 color singlet $E$
with the quantum numbers of a proton --- which we will refer to as a
``diquark'' and a ``triquark'' respectively.  These particles have
ordinary gauge couplings and very small couplings to the first two
generations of quarks but couple strongly to the third
generation. Their masses are proportional to free parameters of the
model.

To suppress flavor-changing neutral currents (FCNC) we rely on a
gauge-mediated supersymmetry breaking scenario \ref\gauge{M. Dine and
W. Fischler, Phys. Lett. {\bf B110} (1982) 227; C. Nappi and B. Ovrut,
Phys. Lett. {\bf B113} (1982) 1751; K. Inoue et al.,
Prog. Theor. Phys. {\bf 67} (1982) 1889; J. Ellis, L. Ibanez and G.G. Ross,
Phys. Lett. {\bf B113} (1982) 283}.  A viable possibility is to append
another sector to the model which breaks supersymmetry and contains
``messenger'' quarks and leptons at $\sim 30$ TeV, as in
\ref\dnns{M. Dine and A. E. Nelson, Phys. Rev. {\bf D48} (1993) 1277,
hep-ph/9303230; M. Dine, A. E. Nelson and Y.  Shirman,
Phys. Rev. {\bf D51} (1995) 1362, hep-ph/9408384; M. Dine, A. E.  Nelson,
Y. Nir and Y. Shirman, Phys. Rev. {\bf D53} (1996) 2658, hep-ph/9507378},
though perhaps a more compelling solution can be found.  Compositeness
effects change the predictions for squark masses; $SU(6)$ relations
imply that the right handed top squark, which contains two preons
carrying $SU(3)$ color, gets a larger soft mass then the left-handed
top squark, which contains one colored and one colorless preon.

Somewhat above the confinement scale it becomes possible to produce
the resonances which are expected in theories with new strong
interactions, which will occur as supermultiplets transforming in
$SU(6)$ representations.  In analogy with QCD we guess that these will
include vector bosons (plus their spin 0 and 1/2 superpartners), with
quantum numbers allowing them to mix with all the ordinary
$SU(3)\times SU(2)\times U(1)$ gauge bosons.  These resonances will
enhance production of the third family quarks and Higgs bosons at very
high energy colliders. We discuss possible experimental signals for
compositeness in section~4.

Some unpleasant features of the model are that we have to give up
grand unification of the ordinary gauge couplings, and that the
leptons have to be put in as a separate sector in order to ensure a
long life for the proton\foot{If all the compositeness scales of this
model are taken higher than $\sim 10^{15}$ GeV as in \topYukawa\ and
\highscale, then we can maintain ordinary
quark/lepton and gauge coupling unification. It is even possible to
unify the new strong interactions with the standard gauge interactions
\ref\sunine{A.E. Nelson, in progress}.  Unfortunately, with such high
confinement scales we would not find any explicit signals for
compositeness.}. Furthermore, we are required to make one dynamical
assumption regarding the effects of confinement on the supersymmetry
breaking mass terms.  Despite these aesthetic drawbacks, we feel this
model is interesting enough to deserve study, as its features are
quite different from most previous ones.  In particular, some
compositeness models must have a much higher compositeness scale in
order to avoid problems with proton decay.  Many have difficulties
generating the observed hierarchy of fermion masses and mixings
without also generating significant flavor-changing neutral currents,
make several percent corrections to precision electroweak predictions,
and/or require dynamical assumptions which are not known to be correct
in any limit.  This model seems to avoid all of these problems.

In the following section we describe a one-generation version
of the model, and then present the full three-generation model 
by  studying a sequence of effective field theories.

\newsec{A model of composite quarks}

Our model is built around the simplest example of an N=1
supersymmetric gauge theory which is known to confine and to not
dynamically break its global symmetries, i.e.  $SU(2)$ with chiral
superfields in six doublets.  This theory has an $SU(6)\times U(1)_R$
global chiral symmetry. By looking for a low energy effective
description of this theory which has the same global anomalies
\cernrep, moduli space of vacua and gauge invariant operators as the 
high energy theory, Seiberg was able to determine \seiberg\
that the correct low energy effective description of this theory
contains a massless gauge-singlet chiral superfield ${\tt\bf M}_{ij},
(i,j=1\ldots6$,) transforming as a ``quindecuplet'' --- a
fifteen-component antisymmetric tensor of the global $SU(6)$ ---
interacting via the effective superpotential
\eqn\superpot{W={\tt Pf}({\tt\bf M})={1\over 6!}\epsilon^{ijklmn}
{\tt\bf M}_{ij}{\tt\bf M}_{kl}{\tt\bf M}_{mn}\ .}

In our model, the dynamics behind preon confinement into quarks will
be 3 such $SU(2)$ supersymmetric gauge theories.  The preons carry
ordinary $SU(3)_c\times SU(2)_w \times U(1)_y$ interactions, which are
embedded in the usual way into an $SU(5)$ subgroup of the $SU(6)$
global symmetries; a ${\bf 6}$ branches to $({\bf 3,1,-1/3})+({\bf
1,2,1/2})+({\bf 1,1,0})$.  The composite fields {\tt\bf M} include the
quark doublet and up-type antiquark, an up-type Higgs doublet, a
diquark, a triquark, and their superpartners.  The effective
superpotential \superpot\ will be responsible for the top, charm and
up quark Yukawa couplings.  In order to obtain masses for the bottom,
strange and down quarks, we will need to include some additional
massive particles, which are doublets under the confining $SU(2)$
groups.

Without further ado, let us list the gauge and global quantum numbers
of all chiral superfields that will appear in the model.  We indicate
the baryon quantum number in order to demonstrate that it is a good
symmetry (aside from the $SU(2)_w$ anomaly).  Note that, as in the
MSSM, $B$ and $L$ conservation need not be explicitly imposed, but can
be accidental symmetries resulting from a combination of discrete
symmetries and renormalizability.  Baryon number can be guaranteed
simply by imposing a discrete unbroken $Z_2$ R symmetry under which
the superpotential changes sign.  The $Z_2$ symmetry we choose need
not guarantee lepton number conservation. Other unbroken discrete
symmetries can be found which would guarantee lepton number
conservation.  In this paper we will simply assume lepton number is
conserved for simplicity, although it would be interesting in future
work to consider the consequences of allowing lepton number violation.
\vbox{\bigskip
{\centerline{\bf Table: Fields and  Symmetries}
\medskip}
{\begintable
Superfield
|$SU(2)_1$|$SU(2)_2$|$SU(2)_3$|$SU(3)_c$|$SU(2)_w$|$U(1)_y$|
$U(1)_B$|$Z_2$\crthick
${\tt\bf d}_1$
| 2       |1        | 1       | 3       |1        |-1/3|-1/6|$+$\cr
${\tt\bf h}_1$
| 2       |1        | 1       | 1       |2        |1/2|1/2|$+$\cr
${\tt\bf n}_1,{\tt\bf N}_1$,${\tt\bf N}'_1$
| 2       |1        | 1       | 1      |1         |0|-1/2  |$-$\cr
${\tt\bf \bar N}_1$,${\tt\bf \bar N}'_1$
| 2       |1        | 1       | 1       |1        |0|1/2   |$+$\crthick
${\tt\bf d}_2$
| 1       |2        | 1       | 3       |1        |-1/3|-1/6 |$+$\cr
${\tt\bf h}_2$
| 1       |2        | 1       | 1       |2        |1/2|1/2  |$+$\cr
${\tt\bf n}_2,{\tt\bf N}_2$,${\tt\bf N}'_2$
| 1       |2        | 1       | 1      |1         |0|-1/2  |$-$\cr
${\tt\bf \bar N}_2$,${\tt\bf \bar N}'_2$
| 1       |2        | 1       | 1       |1        |0|1/2   |$+$\crthick
${\tt\bf d}_3$
| 1       |1        | 2       | 3       |1        |-1/3|-1/6 |$+$\cr
${\tt\bf h}_3$
| 1       |1        | 2       | 1       |2        |1/2|1/2  |$+$\cr
${\tt\bf n}_3,{\tt\bf N}_3$,${\tt\bf N}'_3$
| 1       |1        | 2       | 1      |1         |0|-1/2  |$-$\cr
${\tt\bf \bar N}_3$,${\tt\bf \bar N}'_3$
 | 1       |1        | 2       | 1       |1        |0|1/2  |$+$\crthick
$\bar d_i,\ (i=1,2,3)$
 | 1       |1        | 1       | $\bar3$ |1        |1/3|-1/3|$-$\cr
$\bar H_i$
 | 1       |1        | 1       | 1       |2        |-1/2|0|$+$\cr
$E_i$
 | 1       |1        | 1       | 1       |1        |-1|-1|$-$\cr
$\bar D_i$
 | 1       |1        | 1       | $\bar3$ |1        |1/3|2/3|$+$\cr
$\bar e_i$
 | 1       |1        | 1       | 1       |1        |1|0|$-$\cr
$\ell_i$
 | 1       |1        | 1       | 1       |2        |-1/2|0|$+$
\endtable}}\vfill

As a warmup, we present a one-generation version of the model, in
which the top quark gains a large Yukawa coupling and the bottom quark
receives a smaller one.  Consider a theory with gauge group
$SU(2)\times SU(3)_c\times SU(2)_w \times U(1)_y$, where the first
group factor is the confining gauge group.  As matter content we take
the fields in the table with subscript $3$.  As superpotential take
\eqn\onegensupera{\eqalign{W=&M{\tt\bf \bar N}_3{\tt\bf N}_3+
M'{\tt\bf \bar N}'_3{\tt\bf N}'_3\cr
&+\eta^E{\tt\bf h}_3{\tt\bf h}_3 E_3
+\eta^{H}{\tt\bf h}_3{\tt\bf n}_3{\bar H}_3
+\eta^{D}{\tt\bf d}_3{\tt\bf n}_3\bar D_3\cr
&+\kappa^d{\tt\bf d}_3{\tt\bf \bar N}_3\bar d_3
+\lambda^H{\tt\bf h}_3{\tt\bf N}_3 {\bar H}_3+
\lambda^D{\tt\bf d}_3{\tt\bf N}_3\bar D_3
+\lambda^e\ell_3\bar e_3{\bar H}_3.}}
Below the scale of the massive doublets the effective superpotential
is
\eqn\onegensuperb{\eqalign{W=
&\eta^E{\tt\bf h}_3{\tt\bf h}_3 E_3
+\eta^{H}{\tt\bf h}_3{\tt\bf n}_3{\bar H}_3
+\eta^{D}{\tt\bf d}_3{\tt\bf n}_3\bar D_3\cr
&-{\kappa^d\lambda^H\over M}{\tt\bf d}_3\bar d_3
{\tt\bf h}_3 {\bar H}_3
-{\kappa^d\lambda^D\over M}{\tt\bf d}_3\bar d_3
{\tt\bf d}_3\bar D_3
+\lambda^e\ell_3\bar e_3{\bar H}_3.}}
At the scale $\Lambda$ the $SU(2)_3$ gauge theory becomes strong and
undergoes the confinement discussed above.  The six preons ${\tt\bf
d}_3,{\tt\bf h}_3,{\tt\bf n}_3$ bind into a quindecuplet containing
the quark doublet $q_3\sim{\tt\bf d}_3{\tt\bf h}_3$, the top antiquark
$\bar{u_3}\sim{\tt\bf d}_3{\tt\bf d}_3$, the up-type Higgs boson
$H_3\sim{\tt\bf h}_3{\tt\bf n}_3$, and two new fields $D_3\sim{\tt\bf
d}_3{\tt\bf n}_3$ and $\bar E_3\sim{\tt\bf h}_3{\tt\bf h}_3$.  The
dynamical superpotential \superpot\ is generated, and the resulting
superpotential is
\eqn\onegensuperc{\eqalign{W=&\Lambda\left(\eta^E \bar E_3 E_3
+\eta^{H}H_3\bar H_3 + \eta^{D}D_3\bar D_3\right)\cr
&+\alpha q_3q_3D_3+\beta q_3\bar u_3 H_3+\gamma \bar u_3D_3\bar E_3\cr
&-{\kappa^d\lambda^H}{\Lambda\over M}q_3\bar d_3{\bar H}_3
-{\kappa^d\lambda^D}{\Lambda\over M}\bar u_3\bar d_3\bar D_3
+\lambda^e\ell_3\bar e_3{\bar H}_3}}
where $\alpha\sim\beta\sim\gamma\sim 1$ are introduced to account for
the fact that the $SU(6)$ symmetry which determined the superpotential
\superpot\ has been weakly broken by the gauge and Yukawa couplings.
The fields $D$ and $E$ are massive; let us ignore them for the moment.
The term $\beta q_3\bar u_3 H_3$ is the top-quark Yukawa coupling; it
is of order one.  The bottom quark Yukawa coupling,
${\kappa^d\lambda^H}{\Lambda\over M}q_3\bar d_3{\bar H}_3$, is
naturally less than one, its exact value set by ${\Lambda\over M}$.
(The bottom quark mass also depends on the ratio $\vev{H_3}/\vev{\bar
H_3}$.)  The term $\eta^{H}H_3\bar H_3$ is the $\mu$ term (the
supersymmetric mass for the Higgs bosons) which is naturally of order
$\Lambda$ or smaller.

Thus, for $\Lambda\sim 1$ TeV, $M\sim 1-40$ TeV, the model naturally
generates a large top quark mass, a smaller bottom quark mass, and an
acceptable $\mu$ term.  The mass of the $\tau$ lepton is put in by
hand.  Two new particles $D$ and $E$ are massive and do not much
affect physics near or below $m_Z$.

We now turn to the construction of the full three-generation model.
The superpotential
\eqn\mastersuper{\eqalign{W=&\sum_a\left(M_a{\tt\bf \bar N}_a
{\tt\bf N}_a+ M_a'{\tt\bf \bar N}'_a{\tt\bf N}'_a\right)\cr
&+\sum_{ai}\big(
\eta^E_{ai}{\tt\bf h}_a{\tt\bf h}_a E_i
+\eta^{H}_{ai}{\tt\bf h}_a{\tt\bf n}_a{\bar H}_i
+\eta^{D}_{ai}{\tt\bf d}_a{\tt\bf n}_a\bar D_i\cr
&+\kappa^d_{ai}{\tt\bf d}_a{\tt\bf \bar N}_a\bar d_i
+\lambda^H_{ai}{\tt\bf h}_a{\tt\bf N}_a {\bar H}_i+
\lambda^D_{ai}{\tt\bf d}_a{\tt\bf N}_a\bar D_i
\big)
+\sum_{ijk}\lambda^e_{ijk}\ell_i\bar e_j {\bar H}_k}} 
is the most general gauge invariant renormalizable superpotential
consistent with the global $Z_2\times U(1)_L$ symmetries, and an
additional global symmetry which prevents trilinear couplings for the
${\tt\bf N}'_a,{\tt\bf \bar N}'_a$. (We forbid these latter couplings
because we find they can result in unnacceptable FCNC.  See section
2.3 for a variation in which lepton numbers $-1$, $+1$ are assigned to
${\tt\bf N}'_a, {\tt\bf \bar N}_a'$, respectively, which suppresses
FCNC, and which we hope could explain the lepton mass hierarchy.)

 For convenience, we make field redefinitions so that the $\eta$
coupling matrices are upper triangular and the $\kappa$ matrix is
lower triangular.  All numerical constants except for those describing
the lepton-${\bar H}$ couplings are assumed to be of order 1.

A complete analysis of the low energy physics of this theory follows
in the next section; here we give a brief summary of the roles played
by the various terms in eq.~\mastersuper.  At each compositeness scale
$\Lambda_a$, the fields  $E_a$, $\bar H_a$, and $\bar D_a$ combine with
the composite fields $\bar E_a\sim
{\tt\bf h}_a{\tt\bf h}_a$, $H_a\sim {\tt\bf h}_a{\tt\bf n}_a$, 
$D_a\sim {\tt\bf d}_a{\tt\bf n}_a$ to get
masses of order $\eta_{aa}\Lambda_a$. Off diagonal terms in the $\eta$
matrices will cause these composite fields to mix slightly; the mixing
angles are proportional to ratios of $\Lambda$'s.  
The quark doublets $q_a$ are composite fields ${\tt\bf
d}_a{\tt\bf h}_a$, and the up type antiquarks $\bar u_a$ are composite
fields ${\tt\bf d}_a{\tt\bf d}_a$. The field $\bar
H_3$ will become the down-type Higgs field of the MSSM.  The couplings
of the down quarks to the $\bar H_3$ are generated by graphs involving
tree level ${\tt\bf N, \bar N}$ exchange and the matrices $\kappa^d$
and $\lambda^H$. Similar graphs, with $\lambda^H$ replaced by
$\lambda^D$, will generate couplings of the $\bar D$'s to up and down
antiquarks.   A linear combination of the
composite fields $H_a$ (which is mostly $H_3$) will become the
up-type Higgs. Its superpotential coupling to the composite quarks is
generated dynamically.  The couplings $\lambda^e_{ij3}$ in the last
line will be responsible for lepton masses. 

\subsec{Obtaining the Low Energy Effective Field Theory}

The model is straightforward to analyze provided that all the gauge
and Yukawa couplings are weak at high energies and
$M_a,M'_a\gg\Lambda_a$.  (Another limit, $\Lambda_a\gg M_a$, will be
briefly discussed in section 2.3).  A realistic pattern of quark
masses and mixing emerges when we assume $M_a\sim M_a'$ and take the
three confining $SU(2)$ couplings equal at short distances (with
dynamical scale $\Lambda_0$.) The lepton mass hierarchy is put in by
hand in the superpotential. We ignore the lepton couplings for the
remainder of this section.

The mass hierarchy for the quarks follows from a hierarchy among the
mass terms $M_a$. Each $SU(2)_a$ confines at a scale $\Lambda_a\sim
M_a^{2/3}\Lambda_0^{1/3}$.  Our assumptions, in particular our choice
of four doublets ${\tt\bf N}_a$, ${\tt\bf N}'_a$, ${\tt\bf \bar N}_a$,
${\tt\bf \bar N}'_a$ of approximately equal mass for each confining
group, will lead to the relations for the natural order of magnitude
of quark masses and mixings:
\eqn\quarkmass{\eqalign{
&m_d/m_s\sim\sqrt{m_u/m_c}\sim
\theta_{12}\sim\left({M_2/M_1}\right)^{(1/3)}\cr
&m_s/m_b\sim\sqrt{m_c/m_t}\sim
\theta_{23}\sim\left({M_3/M_2}\right)^{(1/3)}\cr
&\theta_{13}\sim (M_3/M_1)^{1/3}\ .}}
We can choose the $M_a$ such that these are all satisfied to within a
factor of 3 experimentally. 

With such a large hierarchy of scales, a step-by-step top-down
effective field theory analysis is appropriate. We ignore logarithmic
effects from renormalization group running, since these only give
$\CO(1)$ corrections to our results.

\lfm{Step I:}
At energy scales of order $ M_1\sim M_1' (\sim 3\times  10^8$ TeV), we
integrate out ${\tt\bf N}_1$,
${\tt\bf \bar N}_1$, ${\tt\bf \bar N}'_1$, ${\tt\bf N}'_1$, generating
in the effective  superpotential the terms
\eqn\stepone{-\left({1\over M_1}\right)
\sum_{i=1,2,3}\sum_{j=1,2,3}{\tt\bf d}_1 \kappa^d_{1j} \bar d_j
\left( {\tt\bf d}_1\lambda^D_{1i}\bar D_i+
{\tt\bf h}_1\lambda^H_{1i}\bar H_i\right)\ .}
The effective $SU(2)_1$ gauge theory now has six light doublets and will
eventually confine.

\lfm{Step II:}
Below the scale $M_2'\sim M_2(\sim 3\times 10^5$ TeV), we integrate
out ${\tt\bf N}_2$, ${\tt\bf \bar N}_2$, ${\tt\bf \bar N}'_2$, and
${\tt\bf N}'_2$, inducing the superpotential terms
\eqn\steptwo{-\left({1\over M_2}\right)
\sum_{i=1,2,3}\sum_{j=2,3}
{\tt\bf d}_2 \kappa^d_{2j}\bar d_j
\left({\tt\bf d}_2\lambda^D_{2i}\bar D_i+{\tt\bf h}_2
\lambda^H_{2i}\bar H_i
\right)\ .}
Now the $SU(2)_2$ gauge theory also has six light doublets.

\lfm{Step III:} 
Below the  $SU(2)_1$ confinement scale $\Lambda_1(\sim 3 \times
10^4$ TeV), we write down an effective theory for the composite
degrees of freedom $D_1\sim {\tt\bf d}_1{\tt\bf n}_1$, $\bar E_1\sim
{\tt\bf h}_1{\tt\bf h}_1$, $H_1\sim {\tt\bf h}_1{\tt\bf n}_1$,
$q_1\sim{\tt\bf d}_1{\tt\bf h}_1$, and $\bar u_1\sim {\tt\bf
d}_1{\tt\bf d}_1$. The dynamical couplings \superpot\ are written in
terms of these fields as the effective superpotential
\eqn\stepthree{\alpha_1q_1q_1D_1
+\beta_1q_1\bar u_1H_1+\gamma_1D_1\bar u_1 \bar E_1\ } 
where $\alpha$, $\beta$ and $\gamma$ are of order one and are equal up
to small $SU(6)$-breaking effects. For simplicity of presentation, we
will set the dynamically generated $\alpha,\beta, \gamma$ couplings
equal to 1 in eqs (2.10), (2.12), and (2.13). Somewhat below this
scale, couplings in the original superpotential produce mass terms
marrying $\bar E_1$ to $ E_1$, $H_1$ to $\bar H_1$ and $D_1$ to $\bar
D_1$. Only the fields $q_1$ (the up and down quarks) and $\bar u_1$
(the up antiquark) survive to low energies. We integrate out the other
composite fields. The couplings induced in the effective
superpotential for the light fields are:

\eqn\stepthreep{\eqalign{
\sum_{i=1,2,3}\sum_{j,k=2,3}&\Bigg\{- \left({\Lambda_1\over
M_1}\right) \kappa^d_{1i} \bar d_i\left(q_1\lambda^H_{1j}\bar H_j
+\bar u_1\lambda^D_{1j}\bar D_j\right) +\left({1\over M_1}\right)
\left({\lambda^H_{11}\over\eta^H_{11}}
+{\lambda^D_{11}\over\eta^D_{11}}\right)q_1 q_1 \bar u_1\kappa^d_{1i}
\bar d_i\cr &-\left({1\over\Lambda_1}\right) \left[
\left({1\over\eta^H_{11}}\right)q_1\bar u_1 {\tt\bf
h}_j(\eta^H_{j1}{\tt\bf n}_j+\lambda^H_{j1}{\tt\bf N}_j)
+\left({1\over\eta^D_{11}}\right)q_1q_1{\tt\bf d}_j
(\eta^D_{j1}{\tt\bf n}_j+\lambda^D_{j1}{\tt\bf N}_j)\right]\cr
&+\left({1\over\Lambda_1^2}\right) \left({\eta^E_{k1}\over
\eta^E_{11}\eta^D_{11}}\right) \bar u_1{\tt\bf d}_j(\eta^D_{j1}{\tt\bf
n}_j+ \lambda^D_{j1}{\tt\bf N}_j) {\tt\bf h}_k{\tt\bf h}_k\cr &+
\left({1\over M_2\Lambda_1}\right){\tt\bf d}_2\kappa^d_{2j}\bar d_j
\left({\lambda^H_{21}\over\eta^H_{11}}q_1\bar u_1{\tt\bf h}_2+
{\lambda^D_{21}\over\eta^D_{11}}q_1q_1{\tt\bf d}_2\right)\cr &
-\left({1\over M_2\Lambda^2_1}\right)
\left({\lambda^D_{21}\eta^E_{k1}\over\eta^D_{11}\eta^E_{11}}\right)
\bar u_1{\tt\bf d}_2{\tt\bf d}_2 \kappa^d_{2j}\bar d_j {\tt\bf h}_k{\tt\bf
h}_k\Bigg\}\ .\cr}}

\lfm{Step IV:}
$SU(2)_2$ confines at $\Lambda_2 \sim 300$ TeV. Below this scale we
rewrite the theory in terms of the light composite states
$D_2,q_2,H_2,\bar u_2, \bar E_2$, with superpotential couplings
\eqn\stepfour{\alpha_2 q_2q_2D_2+\beta_2q_2\bar
u_2H_2+\gamma_2D_2\bar u_2 \bar E_2\ .} 

Couplings in \mastersuper\ result in masses for $D_2$, $\bar D_2$, $
E_2$, $ \bar E_2$, $ H_2$, $\bar H_2$; integrating them out leads to
superpotential terms
\eqn\stepfourp{\eqalign{
&\sum_{i=1,2,3}\sum_{j=2,3}\Bigg\{ -\left({\Lambda_2\over M_2}\right)
\kappa^d_{2j}\bar d_j
\left(\bar u_2 \lambda^D_{23}\bar D_3+ q_2\lambda^H_{23}\bar H_3\right)
+\left({1\over M_2}\right)\left({\lambda^H_{22}\over\eta^H_{22}}
+{\lambda^D_{22}\over\eta^D_{22}}\right)
q_2q_2\bar u_2\kappa^d_{2j}\bar d_j\cr
&-\left({1\over\Lambda_2}\right)
\left[
\left({1\over\eta^H_{22}}\right)
q_2\bar u_2 {\tt\bf h}_3
\left(\eta^H_{32}{\tt\bf n}_3+\lambda^H_{32}{\tt\bf N}_3\right)
+\left({1\over\eta^D_{22}}\right)
q_2q_2{\tt\bf d}_3
\left(\eta^D_{32}{\tt\bf n}_3+\lambda^D_{32}{\tt\bf N}_3\right)
\right]\cr
&+\left({1\over\Lambda_2^2}\right)
\left({\eta^E_{32}\over\eta^E_{22}\eta^D_{22}}\right)
\bar u_2{\tt\bf d}_3
\left(\eta^D_{32}{\tt\bf n}_3+\lambda^D_{32}{\tt\bf N}_3\right)
{\tt\bf h}_3{\tt\bf h}_3\cr
&+\left({\Lambda_1\over M_1 \Lambda_2}\right)
q_2\kappa^d_{1i}\bar d_i\left(
{\lambda^D_{12}\over\eta^D_{22}}q_2\bar u_1
+{\lambda^H_{12}\over\eta^H_{22}}q_1\bar u_2
\right)
-\left({\Lambda_1\over M_1 \Lambda_2^2}\right)
\left({\lambda^D_{12}\eta^E_{32}\over\eta^E_{22}\eta^D_{22}}\right)
\bar u_1\bar u_2
\kappa_{1i}^d\bar d_i{\tt\bf h}_3{\tt\bf h}_3\cr
&+\left({\Lambda_2\over M_2\Lambda_1}\right)q_1\kappa^d_{2j}\bar d_j
\left({\lambda^H_{21}\over\eta^H_{11}}q_2\bar u_1+
{\lambda^D_{21}\over\eta^D_{11}}q_1\bar u_2\right)
\Bigg\}\ .\cr
}}

\lfm{Step V:}
Below the scale $M_3'\sim M_3(\sim 50$ TeV),  we eliminate
${\tt\bf N}_3$,
${\tt\bf \bar N}_3$, ${\tt\bf \bar N}'_3$, ${\tt\bf N}'_3$,
generating the effective superpotential terms

\eqn\stepfive{\eqalign{
&-\left({1\over M_3}\right)
{\tt\bf d}_3\kappa^d_{33}\bar d_3
\left({\tt\bf d}_3\lambda^D_{33}\bar D_3
+{\tt\bf h}_3\lambda^H_{33}\bar H_3\right)\cr
&+\left({1\over \Lambda_1 M_3}\right)
q_1{\tt\bf d}_3\kappa^d_{33}\bar d_3
\left({\lambda^H_{13}\over\eta^H_{11}}\bar u_1{\tt\bf h}_3
+{\lambda^D_{13}\over\eta^D_{11}}q_1{\tt\bf d}_3\right)
-\left({1\over \Lambda^2_1 M_3}\right)
\left({\eta^E_{31}\lambda^D_{13}\kappa^d_{33}\over
\eta^E_{11}\eta^D_{11}}\right)
\bar u_1\bar d_3{\tt\bf h}_3{\tt\bf h}_3{\tt\bf d}_3{\tt\bf d}_3\cr
&-\left({1\over \Lambda_2 M_3}\right)
q_2{\tt\bf d}_3\kappa^d_{33}\bar d_3
\left({\lambda^H_{23}\over\eta^H_{22}}\bar u_2{\tt\bf h}_3
+{\lambda^D_{23}\over\eta^D_{22}}q_2{\tt\bf d}_3\right)
+\left({1\over \Lambda_2^2 M_3}\right)
\left({\eta^E_{32}\lambda^D_{23}\kappa^d_{33}\over
\eta^E_{22}\eta^D_{22}}\right)
\bar u_2{\tt\bf d}_3{\tt\bf d}_3\bar d_3{\tt\bf h}_3{\tt\bf h}_3\ .
}}

\lfm{Step VI:} For reasons which will be explained in section~3, we
expect soft supersymmetry breaking masses for scalars and gauginos,
which are of order 100-1000 GeV, to be generated at a scale of about
30 TeV.

\lfm{Step VII:} 
$SU(2)_3$ confines at $\sim 1$~TeV. Because the supersymmetry 
breaking masses for the preons are small compared with this scale, we
expect them to have little to no effect on the confining dynamics.  We
will make the assumption at this point that the combination of
confinement and supersymmetry breaking does not give expectation
values to fields carrying color.  (This assumption is discussed in
section 3.3.)  We write down the effective superpotential below this
scale in terms of the light composite and fundamental fields:

\eqn\stepsix{\eqalign{W_{\rm eff}=&\Lambda_3 (\eta^H_{33} H \bar H
+\eta^D_{33} D\bar D+\eta^E_{33}\bar E E)
+\alpha_3q_3q_3D+\beta_3q_3\bar u_3 H+\gamma_3\bar u_3D\bar E\cr
&-\left({\Lambda_3\over\Lambda_2}\right)
\left({\eta^H_{32}\beta_2\over\eta^H_{22}}q_2\bar u_2 H
+{\eta^D_{32}\alpha_2\over\eta^D_{22}}q_2q_2D\right)
-\left({\Lambda_3\over\Lambda_1}\right)
\left({\eta^H_{31}\beta_1\over\eta^H_{11}}q_1\bar u_1 H
+{\eta^D_{31}\alpha_1\over\eta^D_{11}}q_1q_1D\right)\cr
&-\left({\Lambda_3\over M_3}\right)
\kappa^d_{33}\bar d_3\left(\lambda^H_{33}q_3 \bar H
+\lambda^D_{33}\bar u_3\bar D\right)
-\left({\Lambda_2\over M_2}\right)
\left(\kappa^d_{22}\bar d_2+\kappa^d_{23}\bar d_3\right)
\left(q_2\lambda^H_{23}\bar H+\bar u_2 \lambda^D_{23}\bar D\right)\cr
&-\left({\Lambda_1\over M_1}\right)
\left(\kappa^d_{11}\bar d_1
+\kappa^d_{12}\bar d_2+\kappa^d_{13}\bar d_3\right)
\left(q_1\lambda^H_{13}\bar H+\bar u_1 \lambda^D_{13}\bar D\right)\cr
&+\left({\Lambda_3^2\over\Lambda_2^2}\right)
\left({\eta^E_{32}\eta^D_{32}\gamma_2\over\eta^E_{22}
\eta^D_{22}}\right)\bar u_2 D\bar E
+\left({\Lambda_3^2\over\Lambda_1^2}\right)
\left({\eta^E_{31}\eta^D_{31}\gamma_1\over\eta^E_{11}
\eta^D_{11}}\right)\bar u_1 D\bar E\cr
&\quad \quad +{\rm nonrenormalizable\ couplings}}}
We have dropped the ``3'' subscript on the $H$, $D$ and $E$ fields,
since only one linear combination remains of each.  The
nonrenormalizable superpotential couplings are all suppressed by mass
scales of $\Lambda_2$ or higher.  A discussion of observable low
energy effects from effective nonrenormalizable terms, as well as an
explanation for why we choose $\Lambda_3\sim 1$~TeV, can be found in
section~4. A discussion of supersymmetry breaking effects is in
sections 3.3-3.4.

 Below $\sim 1$ TeV the model resembles the minimal supersymmetric
standard model, with the addition of the massive $E$ and $D$
superfields.  The up quark Yukawa couplings to the up-type Higgs boson
are diagonal, with the $i$th generation quark receiving a coupling of
order $\Lambda_3/\Lambda_i\sim (M_3/M_i)^{2/3}$.  The down quark
Yukawa coupling matrix is lower triangular, with the natural size of
the entries in row $i\propto \Lambda_i/M_i\sim (\Lambda_0/M_i)^{1/3}$.
Thus the natural size of Cabibbo-Kobayashi-Maskawa mixing between
families $i$ and $j$ is $\propto {m_d}_i/{m_d}_j$.  This is about what
is seen for the second and third families, and about a factor of five
too small for the first and second families.  There is no specific
requirement on $\tan\beta$, since we can adjust the overall scale of
down-type Yukawa couplings by shifting the $M_a$.  However, since the
top quark Yukawa coupling is a strong interaction effect, we do not
expect that $\tan\beta$ will be much larger than one.  The $D$ and
$E$ couple mainly to third family quarks; their masses, like the $\mu$
parameter $\Lambda_3 \eta_{33}^H$, are undetermined, but cannot be
much above $\Lambda_3$ and certainly can be smaller.  Note that all
$D$ and $\bar D$ couplings to quarks are of the same natural size as
the quark-Higgs boson couplings and are aligned in the same basis,
providing more than adequate suppression of the FCNC generated by $D$
exchange in box diagrams.

 We leave a discussion of electroweak symmetry breaking and
supersymmetry breaking terms for section~3.

\subsec{A minor variation with no strong CP problem}
It is amusing to note that if only the second and third family quarks
are composite, the model naturally predicts a massless up quark, which
could explain the small size of strong CP violation
\ref\masslessup{H. Georgi and I. N. McArthur, Harvard preprint
HUTP-81/A011 (1981) (unpublished); D. B. Kaplan and A. V. Manohar,
 Phys. Rev. Lett. {\bf 56} (1986) 2004; K. Choi, C.W. Kim and
 W.K. Sze, Phys. Rev. Lett. {\bf 61} (1988) 794; K. Choi,
 Nucl. Phys. {\bf B383} (1992) 58; Phys. Lett. {\bf B292}
 (1992) 159, hep-ph/9206247; T. Banks, Y. Nir and N. Seiberg,
 hep-ph/9403203 }.  The down quark mass and the Cabbibo angle need not
 vanish. This variation can be regarded as a limiting case of the
 model described in the preceding section, with
 $\Lambda_1\rightarrow\infty, M_1\rightarrow\infty, \Lambda_1/
 M_1\rightarrow m_d/\langle\bar H\rangle$.

\subsec{When the $\Lambda$'s are large} 
If any or all of the confining $SU(2)$'s become strong at a scale
$\Lambda_a\gg M_a$, the effective theory analysis is very
different. Seiberg has shown that the supersymmetric $SU(2)$ gauge
theory with 8 or 10 massless doublets flows to a superconformally
invariant strongly interacting infrared fixed point (IRFP)
\ref\duality{ N. Seiberg, Nucl. Phys. {\bf B435} (1995) 129, 
hep-th/9411149 }.  We expect this to be approximately the
case for our model as well when $M_a,M'_a\ll\Lambda_a$, although in
the extreme infrared the masses for the doublets ${\tt\bf N}_a$,
${\tt\bf N}'_a$, ${\tt\bf\bar N}_a$, ${\tt\bf\bar N}'_a$ will push the
dynamics away from the fixed point, causing the theory to confine and
produce the same light particles as the limit described in the
preceding section.  However, in this case the theory is strongly
coupled for a long momentum range above the confinement scale, whereas
in the preceding section we assumed weakly coupled descriptions both
above and below the confinement scale.

It is attractive to consider this regime because our superpotential
has many free parameters which would be determined by properties of
the IRFP.  For instance, we can assign lepton number to ${\tt\bf
N}'_a$, ${\tt\bf \bar N}'_a$ and add couplings of leptons to the
preons, of the form $\kappa_{ai}^\ell {\tt\bf h}_a {\tt\bf \bar
N}'_a\ell_i$, to eq.~\mastersuper.  When the theory is approximately
governed by the IRFP over a large energy range, the lepton and $\bar
H$ fields acquire anomalous dimensions of order one.  Such anomalous
dimensions could explain the hierarchy of lepton masses, as well as
the quark masses and mixing angles, as in \ref\democracy{H. Georgi,
A. Nelson and A. Manohar, Phys. Lett. {\bf B126} (1983) 169  }.
However, there are important subtleties involved with this idea, and
it seems we cannot say anything about the theory in this limit without
doing a fair bit of speculation.  We leave this for a future
publication.
 
\newsec{Breaking Supersymmetry and electroweak symmetry}
\subsec{Hidden sector breaking}
It is usually assumed that supersymmetry is spontaneously broken in a
``hidden'' sector, which couples only via supergravity
\ref\sugra{See for example
B. Ovrut and J. Wess, Phys. Lett. {\bf B112} (1982) 347; R. Barbieri,
S. Ferrara and D.V. Nanopoulos, Phys. Lett. {\bf B113} (1982) 219;
J. Ellis and D.V. Nanopoulos, Phys. Lett. {\bf B116}  (1982) 1333;
H.P. Nilles, Nucl. Phys. {\bf B217} (1983) 366; A.H. Chamseddine,
R. Arnowitt and P. Nath, Phys. Rev. Lett. {\bf 49} (1982) 970;
L. Ibanez, Phys. Rev. Lett. {\bf 118} (1982) 73; R. Barbieri,
S. Ferrara and C. Savoy, Phys. Lett. {\bf B119} (1982) 343;
E. Cremmer, P. Fayet and L. Giradello, Phys. Lett. {\bf B120} 
(1983) 41; E. Cremmer, S. Ferrara, L. Girardello and A. Van Proeyen,
Nucl. Phys. {\bf B212} (1983) 413}.  Planck scale physics communicates
supersymmetry breaking to the visible sector, leading to apparent
explicit soft supersymmetry breaking terms.  In order that squark
exchange does not produce excessive flavor changing neutral currents,
it is also usually assumed that the resulting supersymmetry breaking
contribution to scalar masses is universal at the Planck scale. If
squark masses are kept nearly degenerate by an approximate symmetry
(which is broken only by small superpotential couplings), then a
``super GIM'' mechanism prevents large FCNC such as might contribute
to the $K_L-K_S$ mass difference \ref\hlw{ L. Hall, J. Lykken and
S. Weinberg, Phys. Rev. {\bf D27} (1983) 2359}.  However Hall,
Kostelecky and Raby pointed out that the squark mass degeneracy is
violated by renormalization effects below the Planck scale, and so
theories which do not have approximate nonabelian flavor symmetries
for the first two families may have difficulties with FCNC
\ref\fcnc{L.J. Hall, V.A. Kostelecky and S. Raby, Nucl.
Phys. {\bf B267} (1986) 415}.

A way to avoid FCNC without squark degeneracy is to use approximate
abelian symmetries to align the squark masses with the quark masses,
so that, for example, the down and strange squark masses are diagonal
in the same basis as the down and strange quark masses \ref\alignment{
Y. Nir and N. Seiberg, Phys. Lett. {\bf B309} (1993) 337,
hep-ph/9304307}. Note that for the left handed squarks, it is not
possible to align both the up and down squark masses, since the soft
supersymmetry breaking terms are $SU(2)_w$ symmetric. Because of the
small $K_L-K_S$ mass difference, it is phenomenologically necessary to
align the left handed down squark masses rather than the left handed
up squark masses.

In our model, there is no approximate abelian or nonabelian flavor
symmetry for the quarks at any scale, and no reason to expect that the
Planck scale physics which communicates supersymmetry breaking should
respect any such symmetry.  Even if some miraculous mechanism provides
degenerate squark masses at the Planck scale, the first and second
family quarks have strong couplings of very different strengths below
the Planck scale, which will induce substantial (order 1)
nondegeneracy in the renormalized squark masses.

Although nondegenerate, the renormalized squark masses will tend to
align with the quark masses, since the squark mass nondegeneracy is
produced by the same physics responsible for the quark mass
hierarchy. For the left handed squarks, the alignment will be with the
left handed up quarks.  Thus $D-\bar D$ mixing could be suppressed. We
see no way to account for the small size of the $K_L-K_S$ mass
difference, unless the first two family squarks are very heavy ($\sim
5$ TeV).

We believe the experimental absence of large FCNC is strong evidence
that if this model or any similar approach is correct, then
supergravity is not the messenger of supersymmetry breaking.  We must
therefore look well below the Planck scale for the supersymmetry
breaking and the messenger interactions.

\subsec{New mechanisms for supersymmetry breaking in composite models}
The most attractive possibility is that the same dynamics which
produces the composite quarks could also result in supersymmetry
breaking.  Indeed many examples
\ref\iss{K. Intriligator, N. Seiberg and S.H. Shenker,
Phys. Lett. {\bf B342} (1995) 152,
hep-ph/9410203},\sunine,\ref\newmodels{ P. Pouliot, Phys. Lett. {\bf
B367} (1996) 151, hep-th/9510148; P. Pouliot and M.J. Strassler,
Phys. Lett. {\bf B375} (1996) 175, hep-th/9602031; T. Kawano,
YITP-96-5, hep-th/9602035; E. Poppitz, Y. Shadmi and S. P. Trivedi,
EFI-96-15, hep-th/9605113; EFI-96-24, hep-th/9606184; C. Csaki,
L. Randall and W. Skiba, MIT-CTP-2532, hep-th/9605108; C. Csaki,
L. Randall, W. Skiba and R. G. Leigh, MIT-CTP-2543, hep-th/9607021 }
are now known of supersymmetric theories in which gauge boson
confinement, in conjunction with a superpotential, leads to dynamical
supersymmetry breaking \ref\witten{E. Witten, Nucl. Phys. {\bf B202}
(1982) 253}.  Most of these examples involve two or more gauge groups
\newmodels, and a careful analysis of the constraints following from
confinement in one or more of the groups, the superpotential, and
gauge D-terms is required in order to uncover the dynamical
supersymmetry breaking.

 In the limit that all couplings except the confining $SU(2)$'s are
turned off, our theory has a moduli space of supersymmetric ground
states \seiberg.  We have treated the superpotential terms
\mastersuper\ perturbatively, and not found any mechanism whereby
these could induce supersymmetry breaking.  Our model therefore
appears to have the MSSM, without soft supersymmetry breaking terms,
as its low energy limit.  In our analysis so far we have neglected any
dynamical effects involving $SU(3)_c$ and/or $SU(2)_w$ gauge
interactions. Although it is conceivable that nonperturbative effects
involving standard model gauge groups could lead to dynamical
supersymmetry breaking, the supersymmetry breaking scale would surely
be too small \ref\ads{I. Affleck, M. Dine and N. Seiberg,
Nucl. Phys. {\bf B256} (1985) 557}.  We therefore must modify the
model in order to introduce supersymmetry breaking.

\subsec{Gauge mediated visible sector breaking}

We have outlined in the previous section why our model, like all other
viable supersymmetric models, requires the addition of a
``supersymmetry breaking sector''. We have also explained in
section~3.1 why in order to have acceptably small FCNC, supersymmetry
breaking must be communicated by interactions well below
$\Lambda_2\sim 300$~TeV. The possibility which is safest from FCNC is
to have the ordinary $SU(3)\times SU(2)\times U(1)$ interactions
communicate supersymmetry breaking to the squarks and sleptons, since
these interactions are flavor blind. The first two families of squarks
will then naturally have sufficient degeneracy.  Examples of low
energy supersymmetry breaking sectors with gauge mediated
supersymmetry breaking have been constructed and studied elsewhere
\dnns, and shown to be viable, with supersymmetry breaking
communicated at a scale $\sim 30$~TeV. If we append such a sector to
our model, the main effect will be the generation of mass terms for
superpartners carrying $SU(3)\times SU(2)\times U(1)$ quantum numbers,
proportional to their gauge couplings squared.  While it is
straightforward to compute the supersymmetry breaking masses for the
scalar preons at short distances, the supersymmetry breaking masses
for the scalar $(t,b),\bar t,D,H,E$ receive strong corrections from
the strong $SU(2)_3$ dynamics.  The global $SU(6)$ symmetry can be
used to predict the following approximate relations for the
supersymmetry breaking scalar mass terms $\tilde m^2$.
\eqn\scalarmass{\eqalign{ 
&\tilde m_{\bar t}^2=\tilde m_0^2+2 x \tilde
m_d^2,\quad \tilde m_{q}^2=\tilde m_0^2+ x (\tilde m_d^2+\tilde
m_h^2),\cr &\tilde m_{H}^2=\tilde m_0^2+ x \tilde m_h^2,\quad \tilde
m_D^2=\tilde m_0^2+ x \tilde m_d^2,\cr &\tilde m_{\bar E}^2=\tilde
m_0^2+2 x \tilde m_h^2\ , }} 
where $\tilde m_0^2$ and $x$ are an undetermined constants, and
$\tilde m_d^2,\tilde m_h^2$ are the supersymmetry breaking masses for
the preons ${\tt\bf d}_3,{\tt\bf h}_3$, which in a gauge mediated
supersymmetry breaking scenario are expected to equal the
supersymmetry breaking masses for the $\bar d_i$, $\ell_i$ scalars
respectively.  We expect that $x>0$ since it would be surprising for
the lightest squarks to have the heaviest preons.  The masses $\tilde
m_d^2,\tilde m_h^2$ are the masses renormalized at an energy scale
above $\Lambda_3$. $SU(6)$ symmetry guarantees that eq. \scalarmass\
will survive strong renormalization effects below this scale in the
long distance effective theory, although there will be small
corrections from the explicit $SU(6)$ breaking. The large $SU(6)$
symmetric superpotential couplings in the effective theory cause the
parameters $x$ and $\tilde m_0^2$ to be strongly scale dependent, with
$x$ increasing and $\tilde m_0^2$ decreasing at low energy.

However, it is possible that, for example, $\tilde m_{q}^2=\tilde
m_0^2+ x (\tilde m_d^2+\tilde m_h^2)$ is negative at all scales below
$\Lambda_3$, and in this case color would be broken at a high scale.
We make the dynamical assumption that this does not occur.  If our
assumption is wrong, then we must have the messenger scale of
supersymmetry breaking lower than $\Lambda_3$, in which case the
compositeness of the light fields will be irrelevant for supersymmetry
breaking.  While it may be possible to build a gauge-mediated
supersymmetry breaking model with a messenger sector near 1 TeV, it is
likely that $\Lambda_3$ would even then have to be several TeV, making
the model much less interesting for experiment, though no less viable!

\subsec{Electroweak symmetry breaking}

The large top quark Yukawa coupling, in conjunction with soft
supersymmetry breaking terms, can drive electroweak symmetry breaking
\topbreaks. In our model the top Yukawa coupling is also related by
the global $SU(6)$ symmetry to the $D$, $E$ couplings in
\stepsix, and these couplings also renormalize scalar masses.
Note that eq.~\scalarmass\ predicts that, as
renormalization group running causes
$\tilde m_0^2$ to become negative at low energies, the first scalar
mass squared to go negative is $\tilde m_H^2,$ and so the radiative
electroweak symmetry breaking scenario is possible.  In this
regard the model resembles ordinary weakly coupled supersymmetry.

When the messenger scale is larger than $\Lambda_3$, it is interesting
to consider the possibility that even while $\tilde m_{q}^2$ might be
positive, so that color is unbroken, $\tilde m_{H}^2$ might be
negative even at the confinement scale, making a radiative breaking
scenario unnecessary.  In this case the model would more closely
resemble technicolor or topcolor!  Whether this scenario can occur
(and whether color is unbroken) remains an unanswered dynamical
question.

With elementary quarks and leptons, the gauge mediated supersymmetry
breaking scenarios \dnns\ are highly predictive and (so far)
experimentally acceptable, with all supersymmetry breaking masses
determined in terms of only two parameters once the weak scale is
fixed. In our model there are two undetermined strong interaction
coefficients ($x$ and $\tilde m_0^2$) which affect the top and bottom
squark masses and electroweak symmetry breaking. Thus the
uncertainties due to strong $SU(2)_3$ interactions lead to reduced
predictive power in this model, at least until the $D$ and $E$ are
discovered. In particular, it is possible that the soft supersymmetry
breaking mass for $H$ could be {\it larger} at the confinement scale
than the mass for $\bar H$, due to compositeness effects.  Electroweak
symmetry breaking would then have to be due to a large soft
supersymmetry breaking $H$-$\bar H$ scalar mass term, and
$\tan\beta=\langle H\rangle/
\langle\bar H\rangle$ could be less than 1.

\newsec{Experimental tests of quark compositeness}

\subsec{Low energy signals}

First, we consider higher-dimension terms arising from the
superpotential.  The effective superpotential eq.~\stepsix\ contains
higher dimension terms involving the ordinary quarks, but these are
all suppressed by high mass scales. Since they do not give rise to
FCNC at tree level, or violate any symmetries of the standard model,
their effects are uninteresting at low energies.  As we will argue
below, the $D$ and $E$ fields can easily be taken too heavy or too
decoupled to affect low-energy phenomena either at tree-level or
through loops.  All other superpotential terms are present in the
minimal supersymmetric standard model and need no special analysis.

Actually, this is not quite true; there is one other set of operators
we should discuss.  We have prevented baryon number violation in this
model by imposing renormalizability at intermediate energies and a
$Z_2$ symmetry.  However, this does not evade the usual problem of
dimension-five baryon-number-violating operators, which appear in the
superpotential suppressed only by one power of $M_{{\rm Planck}}$.  As
is well known,
\ref\dimfive{S. Weinberg, Phys. Rev. {\bf D26} (1982) 287;
N. Sakai and T. Yanagida, Nucl. Phys. {\bf B197} (1982) 533;
S. Dimopoulos, S. Raby and F. Wilczek, Phys. Lett. {\bf B112} (1982)
133; J. Ellis, D.V. Nanopoulos and S. Rudaz, Nucl. Phys. {\bf B202}
(1982) 43} these operators generically lead to proton decay at far too
high a rate to be consistent with experiment.  Fortunately, in this
{\it and all other low-energy fermion compositeness models}, the
problem is naturally solved: all such operators are suppressed by at
least one factor of the confinement scale divided by $M_{{\rm
Planck}}$.

We next turn to the higher dimension operators in the K\"ahler
potential and those operators involving standard model gauge fields.
We search for effects of compositeness at low energies by doing an
effective Lagrangian analysis. Since the confining interactions do not
break supersymmetry, and since $SU(6)$ is approximately valid at the
confinement scale, we use a supersymmetric $SU(6)$-invariant effective
Lagrangian below the compositeness scale.  (Sub-leading effects due to
soft supersymmetry breaking and $SU(6)$ breaking terms could be
included if desired.)

The most important corrections come from the low compositeness scale
of the third family quarks and up-type Higgs. Since $q_3$, $\bar t$,
$D$, $\bar E,$ and $H$ transform as a ``quindecuplet'' chiral
supermultiplet ${\tt\bf M}_{ij}$, the lowest dimension
nonrenormalizable terms for the composite fields allowed by the global
symmetries are the dimension 6 operators
\eqn\kahlerterm{
\int d^4\theta 
{ \CC_1\over 3\Lambda_3^2}
\left[\Tr\left({\tt\bf M}^\dagger e^{V} {\tt\bf M}e^{V}\right)\right]^2+
{\CC_2\over\Lambda_3^2}\left\{
\Tr\left({\tt\bf M}^\dagger e^{V} {\tt\bf M}e^{V}{\tt\bf M}^\dagger
e^{V} {\tt\bf M}e^{V}\right)-{1\over6}
\left[\Tr\left({\tt\bf M}^\dagger e^{V}{\tt\bf M}e^{V}\right)\right]^2
\right\}}
where $\CC_{1,2}$ are unknown coefficients of order one, and $e^{V}$
contains the $SU(3)\times SU(2)\times U(1)$ gauge interactions
necessary for standard model gauge invariance. \foot{We normalize
${\tt\bf M}_{ij}$ through the kinetic term $\int
d^4\theta\Tr\left({\tt\bf M}^\dagger e^{V} {\tt\bf
M}e^{V}\right)$. Note also that the usual definition of $\Lambda$ in
the compositeness literature is larger than ours by a factor of
$\sqrt{4\pi}$.} These are of course the supersymmetric generalizations
of the familiar current-current interactions.

Loop effects may also induce dimension 6 terms involving ordinary
$SU(3)\times SU(2)\times U(1)$ gauge fields.  If we use the naive
dimensional analysis power counting scheme
\ref\fourpis{A. Manohar and H. Georgi, Nucl. Phys. {\bf B234} (1984) 189},
which estimates the size of terms in an effective Lagrangian by using
perturbation theory with the largest possible self-consistent cutoff
($4\pi\Lambda$), every additional spacetime derivative is associated
with a factor $1/(4\pi\Lambda)$ and every gauge field with a factor
$g/(4 \pi\Lambda)$. Thus we expect $q_{3L},t_R, H$ compositeness to
induce effective operators involving the ordinary gauge fields such as
\eqn\gaugeterma{\int d^4\theta {\CO(g^2/16\pi^2)\over\Lambda_3^2}
\bar D_{\dot\alpha}\left(e^{-V}\bar
W^{\dot\alpha}e^{V}\right)D_\alpha\left(e^{V}W^\alpha e^{-V}\right)+h.c.}
and
\eqn\gaugetermb{\int d^4\theta 
{\CO(g/16\pi^2)\over\Lambda_3^2} {\tt\bf M}^\dagger e^{V}W_\alpha
  e^{-V} D^\alpha \left(e^{V} {\tt\bf M} e^V\right) +h.c.}  
Because the standard model is weakly coupled at the scale $\Lambda$,
these operators can be expected to be unimportant relative to
eq.~\kahlerterm.

Furthermore, since the top quark, charm quark and up quark do not mix
at all, and since the Higgs boson is not discovered and the top quark
is barely studied, all effects observable now or in the near future
involve the bottom quark and the expectation value of the neutral
up-type Higgs boson.

Consider the $SU(4)\times SU(2)_w$ subgroup of $SU(6)$, where
$SU(3)_c$ is a subgroup of $SU(4)$, and note that $(q_3,H)$ transforms
as a ${\bf (4,2)}$ and that the left-handed bottom quark and neutral
up-type Higgs boson both have $I_3=-\half$.  It follows that the 
current-current interaction involving four bottom quarks, four neutral 
Higgs bosons, or two of each, is given by a single irreducible operator 
in the $I=1$ channel whose coefficient is a unique combination of
$\CC_1$ and $\CC_2$ --- we have chosen our normalization of the
$\CC_i$ so that this combination is $\CC_1+\CC_2$.  Thus, all effects
involving these particles are correlated.  This is a remarkable consequence 
of both $SU(6)$ and supersymmetry, and it leads to interesting predictions
below.

In the context of a nonsupersymmetric theory, effects of operators
induced by top quark compositeness were discussed by Georgi et
al.~\ref\topcomp{H. Georgi, L. Kaplan, D. Morin and A. Schenk,
Phys. Rev. {\bf D51} (1995) 3888, hep-ph/9410307 }.  They considered the
effects of dimension-six operators involving the top quark, left
handed bottom quark and gauge bosons.  Their model independent
analysis found the most stringent constraint on left-handed top quark
compositeness came from the possible four-bottom-quark contribution to
$B_d-\bar B_d$ mixing.  Constraints on right-handed composite top
quarks were much weaker.

Similarly, in our model, the term \kahlerterm\ includes the 4-fermi
interaction term
\eqn\fourf{\left( {\CC_1+\CC_2/4\over6 \Lambda_3^2}\right)\bar
q_{3L}\gamma^\mu q_{3L}\bar q_{3L}\gamma_\mu q_{3L} +
\left({\CC_2\over 8\Lambda_3^2}\right)\bar q_{3L}\gamma^\mu\tau_a q_{3L}\bar
q_{3L}\gamma_\mu\tau_a q_{3L}\ .}
Here $q_{3L}\sim (t_L, V_{tb}b_L+V_{ts} s_L+V_{td} d_L)$.  The term
\fourf\ gives a contribution to the $B^0_H-B^0_L$ mass difference of
order
\eqn\masssplit{\Delta m_B\sim(\CC_1+\CC_2) {|V_{td}^2|B_{B_d}f_B^2 m_B\over
18\Lambda_3^2}\ ,}
which for positive $\CC_1+\CC_2$ has opposite sign compared to the
contribution from the standard model.  The value of $|V_{td}|$
extracted from $B$ meson mixing, assuming the standard model, is close
to $0.01$ \ref\alirev{For a review of the relevant $B$ physics see
A. Ali, DESY-96-106, Lectures given at the 20th International Nathiagali
Summer College on Physics and Contemporary Needs, Bhurban, Pakistan,
24 Jun - 11 Jul 1995, hep-ph/9606324}, but unitarity allows values as
small as $.004$, leaving plenty of room for a large non-standard
contribution. Indeed, one can have the observed value of $\Delta m_B$
with acceptable $B_{B_d},f_B$ and $V_{td}$ as long as
\eqn\lowerbound{\Lambda_3>\CO\left( 0.5\ {\rm TeV}
\sqrt{\CC_1+\CC_2}\right)\ .}

One also needs to consider the effects of operators involving the
Higgs and gauge bosons.  We do not expect observable effects from
operators involving the gauge field strength such as those contained
in eq.~\gaugeterma-\gaugetermb, because of the $g/(16\pi^2)$
suppression factors.  A strong bound on the compositeness scale comes
from the operator
\eqn\fhiggs{\left( { \CC_1+\CC_2\over 6\Lambda_3^2}\right) \left(H^\dagger
{i\buildrel \leftrightarrow \over {D^\mu}} H\right)^2}
which is contained in eq.~\kahlerterm.  A general model independent
analysis of the observable effects of gauge invariant dimension six
terms including Higgs bosons was done by Grinstein and Wise
\ref\grwise{B. Grinstein and M. B. Wise,
Phys. Lett. {\bf B265}  (1991) 326}. They found that the only low energy
observable resulting from the dimension 6 operators with 4 Higgs
fields and 2 covariant derivatives is a custodial $SU(2)$ violating
shift in the $W$ and $Z$ masses. Such a shift would affect the $\rho$
parameter by an amount
\eqn\deltarho{(\Delta \rho)_{{JJ}}=-0.020
\left({\sin^4\beta(\CC_1+\CC_2)(1\ {\rm TeV})^2\over \Lambda_3^2}\right)\
,}
where \eqn\sinbeta{\sin\beta={\langle H\rangle\over 175\ {\rm GeV}}\
.}
The constraint on $\Delta \rho$ from precision electroweak analysis
 \ref\pdg{ L. Montanet et al., Phys. Rev. {\bf D50} (1994) 1173 and
 1995 update available on the PDG WWW pages (URL:
 http://pdg.lbl.gov/)} gives
\eqn\deltarhopdg{\Delta\rho=-0.0015\pm0.0019^{+.0012}_{-.0009}\ ,}
where the last numbers reflect the uncertainties due to the unknown
Higgs mass. In a supersymmetric model with a light Higgs, this should
be taken to mean
\eqn\deltarhoobs{\Delta\rho=-0.0024\pm0.0019\ .}

Interesting corrections to the $Z-b-\bar b$ coupling come from
\kahlerterm\ as well, which contains the interactions
\eqn\higgscomp{\eqalign{&
\left({4\CC_1+\CC_2\over12\Lambda_3^2}\right)\bar q_{3L}\gamma_\mu q_{3L}
H^\dagger
i{\buildrel \leftrightarrow \over {D^\mu}} H +
\left({\CC_2\over4\Lambda_3^2}\right)\bar q_{3L} \gamma_\mu\tau_a q_{3L}
H^\dagger
i{\buildrel \leftrightarrow \over {D^\mu}} \tau_a H\cr
&+\left({\CC_1-2\CC_2\over6\Lambda_3^2}\right)
\bar t_R\gamma_\mu t_R H^\dagger
i{\buildrel \leftrightarrow \over {D^\mu}} H  \ .\cr}}
The operator \higgscomp\ can give important corrections to the top and
bottom $Z$ and $W$ vertices, and was not considered by Georgi et al.
The rate for $Z\rightarrow b\bar b$ will differ from the standard
model rate. We find
\eqn\gammazbb{ (\gamma_b)_{{JJ}}\approx0.047\left(
{\sin^2\beta(\CC_1+\CC_2)
(1\ {\rm TeV})^2\over\Lambda_3^2}\right) \ , }
where $\gamma_b$ is defined by \ref\abc{G. Altarelli, R. Barbieri
and F. Caravaglios, Nucl. Phys. {\bf B405} (1993) 3} 
$\Gamma(Z\rightarrow b \bar b)=
\Gamma^0(Z\rightarrow b \bar b)(1+\gamma_b)$, and $\Gamma^0$ is the
standard model rate.
The LEP and SLC experiments currently indicate that \rbanom
\eqn\gammabobs{\gamma_b=0.023\pm0.007\ .}

Comparison of eqs. \gammazbb\ and \deltarho\ shows that
our model predicts
\eqn\rhogammab{(\Delta\rho)_{{JJ}}=
(-0.44)\sin^2\beta(\gamma_b)_{{JJ}}\ .}
The model is potentially consistent with the results
\gammabobs\ and \deltarhoobs.  If we assume the only 
nonstandard contributions to $\gamma_b$ and $\Delta\rho$ come from
compositeness, then for one-sigma consistency with \gammabobs\ and
\deltarhoobs\ we must have
\eqn\tanbeta{\tan\beta<1.3\ .}
The left handed top and bottom squarks give a positive contribution to
$\Delta\rho$, while in our model a positive $\gamma_b$ is correlated
with a negative compositeness contribution to $\Delta\rho$, leading to
a possible cancellation.  For instance, the values
$\Delta\rho=-0.0024$ and $\gamma_b=.02$ can be consistent with
$\sin\beta=1$ if the supersymmetry breaking contribution to the scalar
$q_3$ mass squared is (60 GeV)$^2$, and left-right scalar mixing is
small.

The operators \higgscomp\ also give nonstandard flavor changing
neutral $b-s-Z$ and $b-d-Z$ couplings. Thus if the terms \higgscomp\
account for the nonstandard $\gamma_b$ measurement, the branching
ratios and decay distributions for $b\rightarrow s \ell^+
\ell^-$, $b\rightarrow d \ell^+\ell^-$, and $b\rightarrow s \nu\bar \nu$
should differ by a factor of $\CO(1)$ from their standard
model values \alirev.   Experiments in the next few years will study
these processes in detail.

Note that the nonstandard weak gauge boson couplings are mainly due to
the composite interactions between the Higgs and $q_3, \bar t$. Unlike
the weak gauge bosons, the photon has no Higgs component.  Thus while
the $W$ and $Z$ couplings to $b$ and $t$ quarks receive significant
compositeness corrections, which could be as large as the standard
model one-loop corrections, the effects of compositeness on the
$b\rightarrow s \gamma$ rate are smaller than the standard model
contribution.

The largest nonstandard contribution to the $K_L-K_S$ mass difference
comes from the compositeness of $q_{2L}\sim(c_L, V_{cb} b_L+V_{cs}
s_L+V_{cd} d_L)$, which is acceptably small provided
\eqn\secondbound{\Lambda_2> \CO\left(200\ {\rm TeV}\right)\ .}
The contribution to $K-\bar K$ mixing from the compositeness of $q_3$
is smaller by a factor of $\left[V_{td}V_{ts}m_t/(V_{cd}V_{cs}
m_c)\right]^2$.  The compositeness-induced nonstandard couplings of
$q_{2}$ and $\bar c$ to the weak gauge bosons are negligible.

 Due to the high compositeness scale of the first family of quarks,
there are no significant effects stemming from the compositeness of
$q_{1}, \bar{u}$.

Another possible signal of top quark compositeness comes from the top
quark Yukawa coupling.  In our model the top quark Yukawa coupling is
a nonperturbative effect. A large top Yukawa coupling runs quickly
towards an infrared fixed line, which typically gives in the MSSM
\eqn\approxtopirfp{m_t\sim 200\ {\rm GeV} \sin\beta\ .}
In the MSSM $m_t/\sin\beta$ cannot be more than about 220 GeV since a
large top Yukawa coupling indicates that the MSSM does not remain
weakly coupled at higher energies
\ref\topirfp{C. T. Hill, Phys. Rev. {\bf D24} (1981) 691;
 J. Bagger, S. Dimopoulos and E. Masso, Phys. Rev. Lett. {\bf 55}
(1985) 1450; C.D. Frogatt, I.G. Knowles and R.G. Moorhouse,
Phys. Lett. {\bf B249} (1990) 273, {\bf B298} (1993) 356; S. Dimopoulos,
L. J. Hall and S. Raby, Phys. Rev. Lett. {\bf 68} (1992) 1984;
V. Barger et al., Phys. Rev. Lett. {\bf 68} (1992) 3394; W.A. Bardeen et
al., Nucl. Phys. {\bf B369} (1992) 33}. However in our model no such
bound need be satisfied.

We leave a more comprehensive analysis of the low energy phenomenology
of quark and Higgs compositeness for a future publication.

\subsec{New particles}

The model predicts that massive fields $D$ and $E$ must exist, in
order that the composite fields of the third generation form an
$SU(6)$ representation.  Both are $SU(2)$ singlets but are
electrically charged, and $D$ is colored.  As a result they do not
affect the $\rho$ parameter and other quantities sensitive to $SU(2)$
violation.  The $E$ does not couple to any pair of light fields, but
$D$ does, and its couplings are not flavor diagonal and can violate
the GIM mechanism \ref\gim{S.L. Glashow, J. Iliopoulos and L. Maiani,
Phys. Rev. {\bf D2} (1970) 1285}.  Fortunately for the model, the
GIM mechanism applies for the first two generations, but the $D$ has a
large coupling to the third generation and can contribute to FCNC in
$B$ physics, either through loops or through direct exchange.  In loop
effects, limits from $b\rightarrow s\gamma$ on the scalar diquark are
strongest and are similar to those on charged higgs bosons.  However,
the scalar diquark receives both a supersymmetric and a supersymmetry
breaking contribution to its mass. The supersymmetry breaking
contribution could easily be larger than 500 GeV.  Furthermore nothing
in the model prevents us from giving the $D$ field a large
supersymmetry preserving mass.  We should therefore view $b\rightarrow
s\gamma$ as a constraint which forces $m_D$ to be large compared with
$m_W$ but moderate compared with the compositeness scale; as such it
does not test the model. Exchange of $D$ fields can induce dimension
five and six terms which can contribute to FCNC's, but given the
$b\rightarrow s\gamma$ constraint these are always subleading to the
standard model contributions.  There are no significant limits on the
$D$ and $E$ fermions or the $E$ scalar beyond the obvious ones from
collider searches.  Note that although the $D$ and $E$ have the gauge
quantum numbers of down quarks and positrons, they are forbidden to
mix with them by baryon number conservation; as a result, there will
be no violations of unitarity in the CKM matrix.

Of course, nothing would substitute for the direct discovery at
colliders of these supermultiplets.  Generically speaking, the main
decay mode of the fermionic $D$, (the ``diquarkino''), which has odd
R-parity, is to two third family quarks and a standard model gaugino,
while the diquark decays mainly to two third family quarks. The
triquark, with even R parity, primarily decays to 3 third family
quarks, and the scalar $E$ (the ``trisquark'') decays to 3 third
family quarks and a gaugino.  Again, the absence of mixing of $D$ and
$E$ with down quarks and positrons distinguishes the decays of these
particles from similar particles in many other models.  Still, the
specific decay modes depend in detail on the masses, both
supersymmetry preserving and breaking, of these and other fields, and
we do not have enough constraints on these masses to make definite
predictions for their decay signatures.

\subsec{Squark masses}

The standard predictions of a gauge-mediated supersymmetry breaking
model with messenger quarks and leptons \dnns\ will apply to all
fundamental particles and those which are composite above the
supersymmetry breaking scale.  In particular, all such quarks have
roughly the same mass, with the $SU(2)_w$ doublet squarks of the first
two generations being slightly heavier than the $SU(2)_w$ singlets.
However, the low energy composite fields satisfy eq.~\scalarmass.  We
expect $x\sim 1$ and $m_0^2<0$ at low energy in order that
electroweak symmetry be broken.  This will then lead to the prediction
that $\tilde m_{\bar t}$ is greater than $\tilde m_{q}$ by a
substantial amount, of order (very roughly!) $40\%$.  The $D$ and $E$
fields may have large supersymmetry preserving masses, but were their
soft supersymmetry breaking masses to be measured, a number of simple
relations, such as $\tilde m_{\bar t}^2+\tilde m_{E}^2\approx 2\tilde
m_{q}^2$, would be strong tests of $SU(6)$.

\subsec{Signals at Multi-TeV Colliders}

The most dramatic signals of quark compositeness could be seen in
collisions at energies above the scale $\Lambda_3$. Here the particle
spectrum is expected to include a multitude of resonances, and the
form factors for the couplings of top quarks, bottom quarks and weak
gauge bosons will differ from their standard model values. If QCD is a
good guide, the resonance region is well above the scale $\Lambda_3$,
by a factor of $\sim 3-10$, and since we expect $\Lambda_3=1-3$~TeV
these would probably be out of reach for LHC and any forseeable lepton
collider. Still it is interesting to examine the likely high energy
signals of the new strong interactions. We expect a huge number of
resonances with quite exotic quantum numbers (color sextets, weak
triplets, charge 2, high spins etc.), but since these will probably
have mass of $\sim 3-30$ TeV they could be out of experimental reach
for the foreseeable future.

The resonances  likely to have the largest effects on  high energy
phenomenology are in a massive vector supermultiplet, with the 
quantum numbers of a  35-plet plus a singlet under the  global 
$SU(6)$. These have ordinary spins $1, 1/2,$ and $0$.  Their ordinary
$SU(3)_c\times SU(2)_w\times U(1)$ quantum numbers are
\eqn\resonance{\eqalign{&({\bf 8},{\bf 1},{\bf 0})+
({\bf 1},{\bf 3},{\bf 0}) + 3 ({\bf 1},{\bf 1},{\bf 0})+
({\bf 3},{\bf 1},{\bf -1/3})+({\bf \bar 3},{\bf 1},{\bf 1/3})\cr&
+({\bf 1},{\bf 2},{\bf 1/2})+({\bf 1},{\bf 2},{\bf -1/2})
+({\bf 3},{\bf 2},{\bf -5/6})+({\bf \bar 3},{\bf 2},{\bf 5/6})\ .}}
Their masses might well be too large to have effects at LHC,  but it is
worth asking how one could observe them  if they are on the light
side.  Perhaps the largest effects at LHC could come from the heavy 
color-octet spin-one resonance, the analogue of the $\rho$-meson,
which mixes with the gluon and couples most strongly to the third
generation quarks.  Potentially it could show up in the channels
\eqn\lhceffect{g g\rightarrow t \bar t,\quad gg\rightarrow b \bar b
,\quad gg\rightarrow gg\ .}
Higgs boson and electroweak gauge boson production could also be
enhanced through some of the other resonances and might be visible at
LHC or at a lepton collider.

Clearly a more comprehensive study is needed here, which we leave for
future work.

\newsec{Conclusions}

We have presented a supersymmetric quindecuplet model, using with
considerable modification the mechanisms of \topYukawa\ and
\highscale, in which certain spin 0 and ${1\over 2}$ particles of the
supersymmetric standard model are composite.  Our dynamical analysis
relies heavily on the work of Seiberg \seiberg.  The quark mass
hierarchy is explained as a hierarchy of confinement scales, with the
compositeness scale of the third generation at $1-3$ TeV.  The up-type
Yukawa couplings are generated dynamically, the down-type Yukawas by
exchange of massive fields and confinement.  If gauge-mediated
supersymmetry breaking is used, flavor-changing neutral currents are
suppressed without fine-tuning.  The model has two new supermultiplets
below a TeV, and a slew of resonances well above a TeV, which couple
predominantly to the third generation.  An approximate global $SU(6)$
symmetry and supersymmetry assure that confinement-scale effects on
the $\rho$ parameter, $Z\rightarrow b\bar b$, $b\rightarrow
s\ell^+\ell^-$, $b\rightarrow d\ell^+\ell^-$, $b\rightarrow s \nu\bar\nu$, 
$b\rightarrow d \nu\bar\nu$ and $B-\bar B$ mixing
are determined by $\tan\beta$ and a single unknown coefficient.  The
relation eq.~\rhogammab\ is particularly unusual and also is
phenomenologically interesting given present constraints on $\rho$ and
$R_b$.  Among the predictions which are probably generic to
low-energy supersymmetric compositeness models are that the usual
relations for soft supersymmetry-breaking masses are significantly
modified by compositeness effects and the problem of dimension-five
baryon-number-violating operators is eliminated.

  While unlikely to be the full story, especially as the lepton mass
hierarchy is unexplained and supersymmetry breaking requires a
separate sector, this quindecuplet model has many interesting and new
elements.  Its ability to avoid many of the classic problems of
compositeness models is remarkable.  Could this be a sign that a
strongly coupled supersymmetric gauge theory is indeed the missing
piece of the phenomenological puzzle?

\medskip

\centerline{\bf Acknowledgements}

 A.E.N was supported in part by the Department of Energy under grant
\#DE-FG06-91ER40614.  M.J.S was supported in part by the Department 
of Energy under grant \#DE--FG05--90ER40559.  We had useful
conversations with Don Finnell, Ian Hinchliffe, David Kaplan, Chris
Kolda and John Terning. M.J.S would like to thank the theory group of
the University of Washington for their extensive hospitality.  We
would also like to thank the Aspen Center for Physics for hospitality
during the completion of this work.
\listrefs
\bye
\end